# Machine learning enables experimental access to photon-by-photon arrival times in scintillation detectors

Yuya Onishi[1,2,3,†], Ryosuke Ota[1], Fumio Hashimoto[4], Kibo Ote[1], Go Akamatsu[3], Hideaki Tashima[3], and Taiga Yamaya[3,2,†]

[1] Central Research Laboratory, Hamamatsu Photonics K.K.

[2] Graduate School of Science and Engineering, Chiba University

[3] Department of Advanced Nuclear Medicine Sciences, National Institutes for Quantum Science and Technology

[4] J. Crayton Pruitt Family Department of Biomedical Engineering, University of Florida

[†]Corresponding authors: yuya.onishi@hpk.co.jp, yamaya.taiga@qst.go.jp

## Abstract

Scintillation detectors with excellent timing resolution enable more precise localization of radiation sources in positron emission tomography, leading to substantial improvements in diagnostic capability for diseases such as cancer and dementia. At the extreme timing precision required for such applications at the picosecond scale, detector performance is governed by the microscopic dynamics of scintillation photons generated within the detector and their subsequent detection processes. However, detector signals have conventionally been treated only as collective responses of many photons due to structural constraints inherent to photodetectors. In this study, we overcome this fundamental limitation using deep learning, enabling direct access to the timing information of individual photons. The proposed method estimates photon-by-photon arrival times directly from detector waveforms without requiring any modification to the detector structure; the method operates on an event-by-event basis without ground-truth labels by integrating an unsupervised learning framework with a physically informed detector-response model. Through comprehensive validation combining Monte Carlo simulation and experimental measurements across various detector configurations, we experimentally demonstrate improved timing resolution, visualized depth-of-interaction-dependent photon transport, and classified Cherenkov and scintillation photons based on the estimated photon-level timing information using a unified deep learning-based framework. These results provide experimental access to photon dynamics, bridging the gap between theoretical modeling and experimental observation, and they open a new data-driven pathway for discovery in detector physics and optimization.

Radiation detection is a key enabling technology that supports a wide range of fields, spanning from fundamental science to societal applications, including particle physics, space science, environmental monitoring, and medical diagnostics and therapy. As the demands of these applications continue to grow, radiation detectors are required not only to detect extremely low levels of radiation, but also to precisely determine the energy, interaction position, and arrival time of incident particles or photons[1]. Among the various radiation detection technologies, scintillation detectors have been widely adopted because of their high detection efficiency, fast response, high sensitivity to most particles, and broad energy applicability.

A prominent example in which scintillation detectors have reached a mature stage of societal implementation is positron emission tomography (PET) in medical imaging. PET is a functional imaging modality that visualizes physiological and metabolic processes in the body using disease-specific radiotracers, providing high diagnostic accuracy for diseases such as cancer and neurodegenerative disorders. Unlike conventional imaging techniques that primarily capture anatomical structures, PET provides functional information at the cellular and molecular levels, allowing the detection of lesions that may not yet be visible morphologically[2]. A positron emitted from the injected tracer annihilates with an electron, producing a pair of gamma rays that are emitted nearly simultaneously in opposite directions and detected in coincidence to localize the origin of the event (Fig. 1a). However, a fundamental limitation of conventional PET scanners is the relatively low detection efficiency of coincidence events, resulting in limited counting statistics and high noise levels in reconstructed images. To address these limitations, time-of-flight (TOF) PET has been introduced as an effective approach as it exploits the slight difference in arrival times of the two gamma-rays to more precisely constrain the annihilation position. By reducing spatial uncertainty, TOF information suppresses noise propagation and improves the signal-to-noise ratio of reconstructed images, significantly contributing to improved diagnostic capabilities[3-6]. Accordingly, TOF capability has become an essential technology in PET systems, with ongoing efforts striving to achieve ever finer timing resolution, which directly impacts diagnostic ability.

To achieve further improvements in timing resolution, it is indispensable to measure the arrival time of extremely fast optical signals with high precision. In PET detectors, incident radiation is converted into scintillation photons within a scintillator crystal, which are subsequently read out as electrical signals by photodetectors. Advances across the detection chain, including scintillator materials, photodetectors such as photomultiplier tubes (PMTs) and silicon photomultipliers (SiPMs), and readout electronics, have led to substantial improvements in timing performance[3-6]. Consequently, timing resolutions on the order of 200 ps have been achieved in state-of-the-art systems[7-9]. As timing performance is pushed toward ever higher precision, the temporal characteristics of scintillation detectors become increasingly dominated by subtle time fluctuations arising from internal physical processes, including the emission, transport, and detection of scintillation photons. Although timing performance has steadily improved through progress in material development and system-level optimization, conventional detectors have typically evaluated performance using macroscopic observables such as coincidence timing resolution (CTR) and light yield. Consequently, the internal photon dynamics that give rise to these observables are often treated as a black box, and optimization has largely relied on empirical tuning rather than direct insight into photon behavior within the detector. On the other hand, the microscopic behavior of scintillation photons has been extensively investigated through theoretical studies and Monte Carlo simulations[6,10], which have provided important insights into photon transport and detection processes[11-14], as well as more extensions incorporating high-aspect-ratio crystals[15], depth-of-interaction (DOI) effects[16-17], and prompt photon contributions[18-20]. In addition, these studies have shown that exploiting multiple scintillation photon timing could, in principle, approach the fundamental limits of timing resolution as described by the Cramér–Rao lower bound. However, these results have remained inherently constrained within analytical and simulation frameworks for many years, as existing detectors do not provide direct access to the individual timing information of scintillation photons. Furthermore, many aspects of detector response and photon interactions, e.g. non-uniformity of scintillator characteristics, surface finish quality, inhomogeneous reflector wrapping and optical glue, and misalignment between scintillators and photodetectors, are often idealized in simulations. The experimental validation of scintillation photon dynamics remains largely unexplored, highlighting the critical need for approaches that enable direct access on a detector-by-detector basis because there are no two exactly identical detectors in practice.

In this study, we aim to computationally estimate and analyze the photon-level timing inside detectors without relying on any specialized devices as we work towards high-precision TOF measurements. The arrival time information of an individual scintillation photon quantized by the proposed method is referred to as a photon timestamp in this study. The proposed approach extracts latent photon timestamps directly from analog signal waveforms by integrating machine learning with physically informed detector-response modeling (Fig. 1a). In general, the detector waveform can be described as the convolution of scintillation kinetics with the SiPM response function, combined with additional electronic noise[6]. The central idea of our approach is that individual scintillation photons can be quantized by removing the SiPM degradation factors from the measured waveform within a deep-learning framework. An advantage of this method is that it does not require ground-truth photon timestamps, which are inaccessible under experimental conditions. Consequently, the method operates as an unsupervised learning framework in which physically motivated response models are embedded directly into the training process. In recent years, deep learning-based signal-processing techniques have been increasingly applied to detector applications[21-23] and they have demonstrated significant CTR improvements over conventional signal-processing methods[24-29]. However, most existing approaches rely on supervised learning and require labeled data obtained from either simulations or highly controlled measurements. Such labels are typically generated under specific modeling assumptions and may not faithfully represent the underlying physical processes, especially when the internal stochastic dynamics of photon generation and detection are not directly observable. Consequently, in scenarios involving unknown or previously uncharacterized physical phenomena, reliance on predefined ground-truth labels constitutes a fundamental limitation. By contrast, our proposed approach eliminates the need for predefined labels and instead enables direct experimental access to event-by-event scintillation and/or Cherenkov photons. This capability provides a new pathway toward understanding the microscopic origins of timing performance and its fundamental limits in scintillation detectors. Such an understanding of photon behavior not only advances the timing performance, the physical modeling of scintillation detection, and new physics observation, but also lays the groundwork for the rational design and optimization of next-generation detectors.

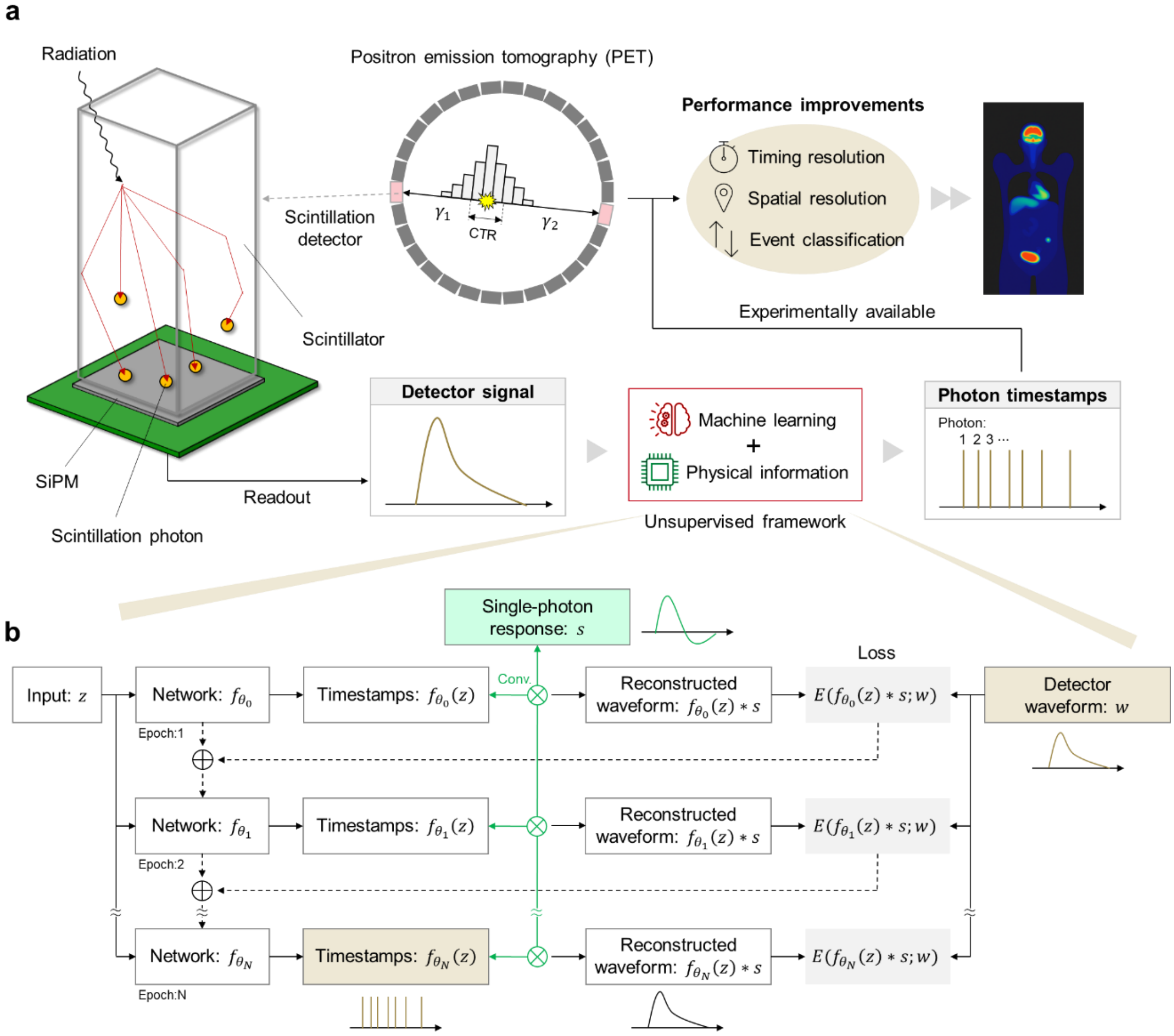

**Fig. 1 | Concept of unsupervised learning-based photon timestamp estimation inside a scintillation detector. a**, Schematic flow chart of this study. A large number of scintillation photons are generated inside the scintillator following radiation interaction, while the detector output is conventionally measured as an analog waveform. In the proposed method, machine learning with physically-available detector information is used to estimate photon timestamps directly from the analog signal. **b**, Overview of the proposed framework. A neural network maps photon timestamps from detector waveforms on an event-by-event basis by explicitly incorporating the single-photon response of the SiPM into a network model, eliminating the need for a prior training dataset.

## Results

We introduced an unsupervised learning-based photon timestamp estimation method for visualizing photon behavior and enhancing timing and spatial resolutions as well as scintillation/Cherenkov event classification performance in scintillation detectors. An overview of the proposed framework is given in Fig. 1b. This framework uses a neural network to map photon timestamps from detector waveforms on an event-by-event basis by explicitly incorporating the single-photon response of the SiPM into a network model, eliminating the need for a prior training dataset. A detailed description of the methodology and network architecture is provided in Methods. To evaluate the validity and effectiveness of the proposed method, we first make a preliminary evaluation using Monte Carlo simulation data. Then, a comprehensive evaluation is made across a diverse range of tasks that have conventionally been investigated independently for timing performance improvement. Lutetium yttrium oxyorthosilicate (LYSO) and bismuth germanate (BGO) scintillators are used for this evaluation; the latter one is considered to be a hybrid Cherenkov/scintillation material. These evaluations include CTR estimation, DOI-dependent photon arrival statistics, and discrimination between Cherenkov and scintillation photons, and they are all based on photon timestamps estimated from experimental data.

### Monte Carlo simulation

We first evaluated the proposed method using Monte Carlo simulation data. The simulation setup implemented in Geant4[30] is illustrated in Fig. 2a. A $3 \times 3 \times 10\,\mathrm{mm}^3$ LYSO crystal was irradiated with 511-keV $\gamma$-rays, and all optical photons generated by $\gamma$-ray interactions within the scintillator were tracked to obtain ground-truth photon arrival times. The information of these timestamps was subsequently converted into analog detector waveforms by applying the real single-photon response of the SiPM and electronic noise (see the section "Monte Carlo simulation setup" for details). Fig. 2b shows representative examples comparing the true waveforms and photon timestamps with those estimated by the proposed method. The reconstructed waveforms are visually similar to the true waveforms, indicating that the detector signal is accurately reproduced within the network model. Importantly, the estimated photon timestamps exhibit a consistent temporal structure with the ground-truth, demonstrating that the proposed framework can simultaneously reconstruct detector signals while quantizing the corresponding photon timing.

To quantitatively assess estimation accuracy, we analyzed the differences between the true and estimated photon timestamps. The timing bias and the corresponding standard deviation are shown in Fig. 2c. Although the results depend on the regularization parameters in the loss function, incorporating regularization or applying time-dependent weighting effectively reduces the timing bias compared with a simple mean squared error. With optimal parameter settings, the timing bias is suppressed to within approximately 10 ps. The timing standard deviation is larger for early photons but gradually decreases to approximately 35 ps for later photons. This is because early photons stronger overlap with the rising edge of the detector signal and increases their susceptibility to electronic noise, making their recovery intrinsically more challenging. Nevertheless, when comparing averaged arrival times over multiple photons, the discrepancy between true and estimated timestamps is further reduced. Although the timestamps are not perfectly estimated with zero error, we consider the achieved temporal accuracy acceptable when taking into account the timing resolution of current detector systems[7-9], indicating that the estimated timestamps are sufficiently precise for practical applications. In addition to timing accuracy, we evaluated the fidelity of total photon count. Fig. 2d shows the distribution of differences between the true and estimated number of photons per event. A Gaussian fit to this distribution yields a peak position of 0 photon equivalent (p.e.) and a standard deviation of 3 p.e., indicating a remarkably small deviation corresponding to 1.3%. This result indicates that the proposed method simultaneously preserves temporal information and count-related fidelity at the photon level.

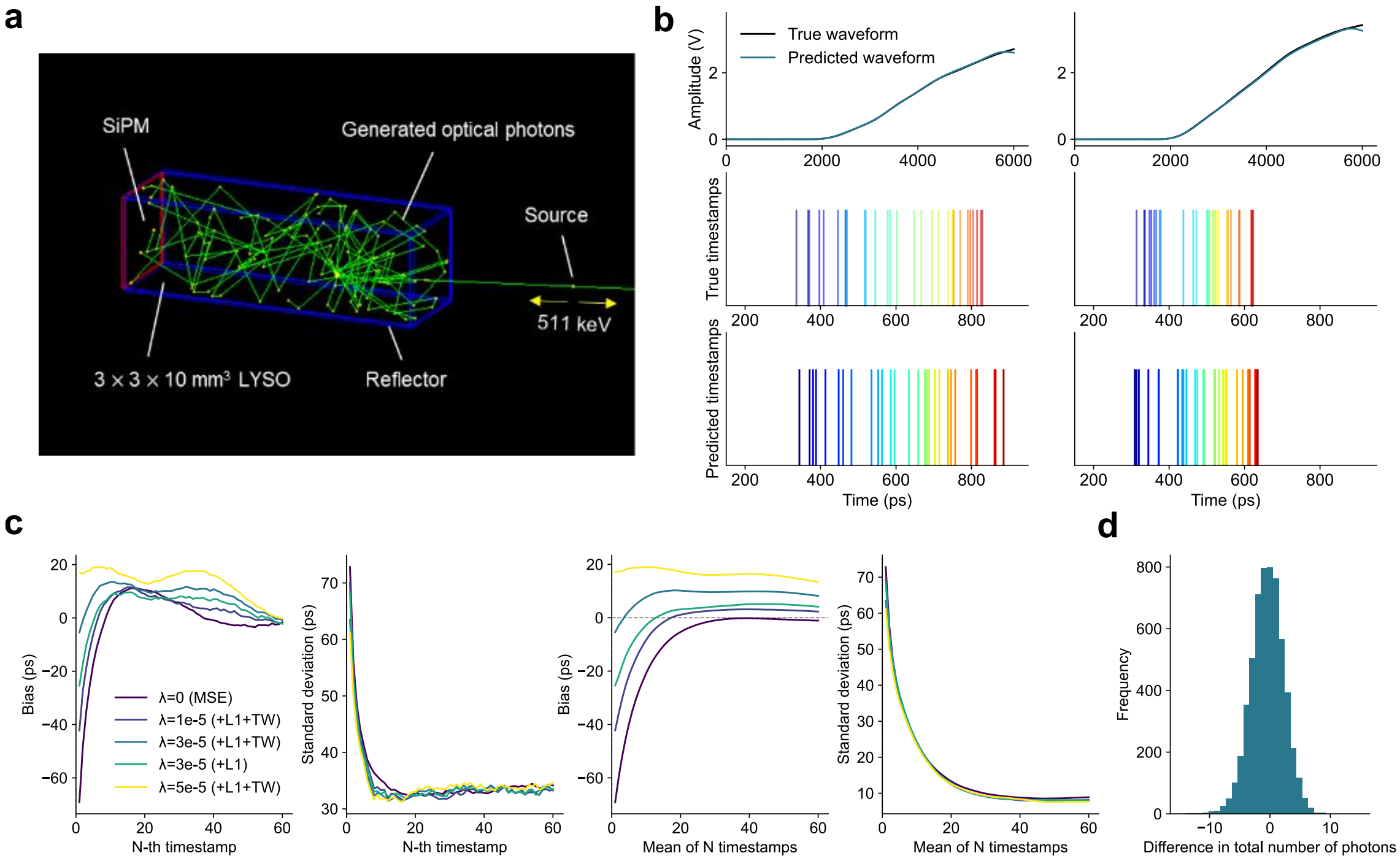

**Fig. 2 | Quantitative comparison between true timestamps generated by Monte Carlo simulation and those estimated by the proposed method.** **a**, Schematic overview of the Geant4-based Monte Carlo simulation setup. This figure is created with fewer photons than the actual simulation conditions. **b**, Representative examples of the true waveform and photon timestamps, together with the corresponding reconstructed waveform and estimated timestamps. **c**, Timing bias and standard deviation calculated from the time differences between the true and predicted timestamps at individual photon and averaged photon timings. **d**, Distribution of the difference between the true and estimated numbers of photons.

**Timing resolution estimated from the multiple timestamps**

After obtaining the simulation results, we experimentally evaluated the timing performance of the estimated timestamps using analog detector signals acquired with a coincidence setup consisting of two 3 × 3 × 10 mm$^3$ LYSO detectors coupled to SiPMs, as shown in Fig. 3 (see the Section "Experimental setup for CTR evaluation" for details). The CTR obtained directly from the analog waveforms reaches 116.3 ± 0.3 ps full width at half maximum (FWHM) when an optimal timing pick-off threshold and an applied voltage of 66 V were set (Fig. 4a). Fig. 4b shows the CTR behavior estimated from photon timestamps estimated by applying the proposed method. Compared with using individual photon timestamps alone (blue curve), incorporating multiple photons results in a clear improvement in CTR. When arrival times are simply averaged over multiple photons, the contribution from later photons whose timing information is degraded becomes dominant, leading to a deterioration in CTR (black curve). In contrast, maximum likelihood-based time estimation (MLTE) appropriately weights early photons (red curve), which carry higher timing information content, and MLTE suppresses the influence of later photons (see the section "Timing estimator using multiple timestamps" for details). As a result, this approach effectively mitigates CTR degradation and enables convergence toward an optimal value, with the best achieved CTR reaching 109.7 ± 0.8 ps FWHM. The observed behavior is consistent with theoretical expectations and previous studies based on simulation and analytical models[11-14].

Furthermore, the estimated photon timing information preserves sensitivity to spatial variations of the positron source (Fig. 4c). In this measurement, the source position was not physically displaced; rather, a virtual offset of 10 mm was introduced by digitally shifting one of the coincidence waveforms, resulting in a corresponding shift of the time-difference distribution. The measured peak displacement of 66.2 ps quantitatively agrees with this imposed offset, confirming that the estimated timestamps accurately capture source position-dependent timing information.

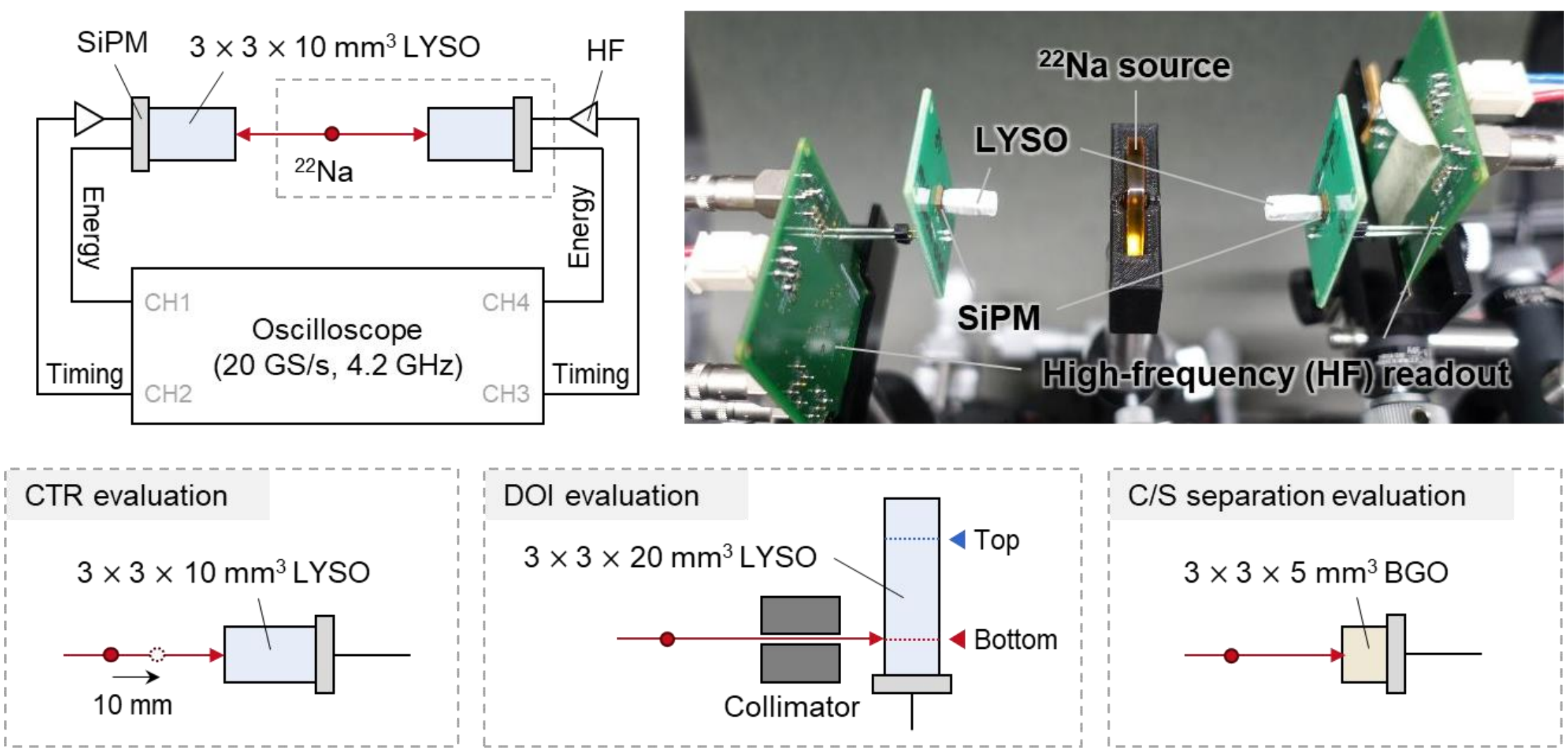

**Fig. 3 | Experimental setups using various scintillation detectors.** Schematic diagrams and a photograph of the

coincidence timing measurement setup consisting of two scintillation detectors irradiated with a $^{22}Na$ source. The detector configuration enclosed by the dashed line box is interchangeable and corresponds to the specific detector configurations shown at the bottom. The detector configurations implemented for each evaluation are as follows: coincidence time resolution (CTR) evaluation using a LYSO crystal with a thickness of 10 mm; depth-of-interaction (DOI) evaluation using a LYSO crystal with a thickness of 20 mm and collimated irradiation at the top and bottom interaction positions; and Cherenkov and scintillation (C/S) separation evaluation using a BGO crystal with a thickness of 5 mm.

---

**Photon arrival time distribution on different interaction positions**

We further investigated the DOI dependence of the estimated photon arrival time distributions. Fig. 3 illustrates the experimental setup used for this evaluation, in which a $3 \times 3 \times 20\ mm^3$ LYSO crystal was irradiated with a collimated $^{22}Na$ source to selectively induce interactions near the top or bottom of the crystal. By constraining the interaction position along the crystal depth, photon arrival time distributions corresponding to different DOI positions were obtained (see the section "Experimental setup for DOI evaluation" for details).

Fig. 4d shows the normalized arrival time distributions for events occurring near the top and bottom regions of the crystal. A shift in the photon arrival time distribution is observed between the two DOI positions, reflecting the difference in optical photon propagation time to the SiPM. For early-arriving photons, bottom events consistently exhibit earlier arrival times than top events, reflecting the shorter optical path length for interactions occurring close to the SiPM. In contrast, this relationship reverses for later photons: bottom events show delayed arrival times compared with top events. This behavior arises because bottom events include optical photons traveling along the longest and most diverse propagation paths due to multiple internal reflections, resulting in a larger photon transfer time spread. Consequently, bottom events exhibit a broader arrival-time distribution than top events, as evident from the overlapping distributions. Importantly, this difference is also directly confirmed using the estimated photon timestamps, which show that top events yield consistently better timing resolution than bottom events. The observed DOI-dependent trends are consistent with a previous study based on an analytical model, further supporting the physical fidelity of the estimated timestamps[16].

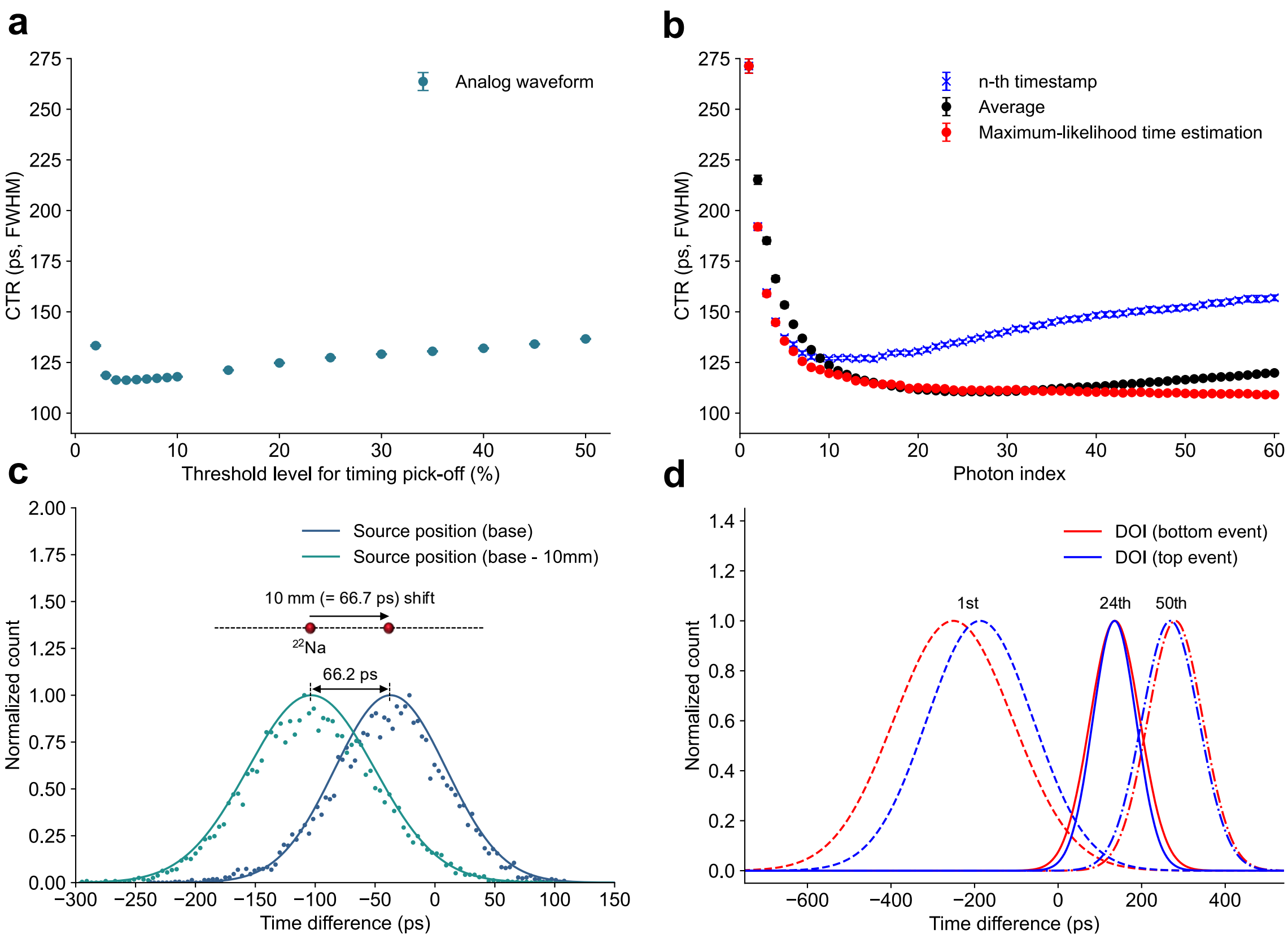

**Fig. 4 | Experimental results of timing performance and DOI-dependent photon arrival time distributions using the estimated photon timestamps. a**, CTR obtained directly from the analog waveforms for different timing pick-off threshold settings. **b**, CTR estimated from the estimated photon timestamps as a function of the photon index. Curves are shown for individual photon timestamps (n-th timestamp), simple averaging and maximum-likelihood time estimation (MLTE) over multiple photons. **c**, Time-difference distributions for two source positions, exhibiting a measured peak shift of 66.2 ps, close to the true value of 66.7 ps. **d**, Normalized photon arrival time distributions for top and bottom interaction positions. The three distributions, shown from left to right, correspond to the 1st, 24th, and 50th detected photons, respectively.

## Separation of Cherenkov and scintillation photons

To evaluate separation performance between Cherenkov and scintillation photons, we acquired coincidence waveforms using a detector configuration in which a 5-mm-thick BGO crystal was placed face-to-face with a LYSO-based reference detector (see the section "Experimental setup for BGO evaluation" for details). From the obtained waveforms, a time-difference histogram was constructed by subtracting the timing pick-off value of the reference detector from that of the BGO detector. The resulting asymmetric time-difference histogram was fitted with a composite model consisting of a Gaussian function representing the Cherenkov contribution and an exponentially modified Gaussian (EMG) function accounting for scintillation photons (Fig. 5e, left). For subsequent analyses, time windows were defined on the time-difference histogram to classify events into fast ($T_{BGO}$-$T_{ref} \leq -0.05$ ns) and slow ($T_{BGO}$-$T_{ref} \geq 0.3$ ns) components. The numbers of fast and slow events were adjusted to be equal.

For both fast and slow events, Fig. 5a shows the distributions of time differences between a specific photon and the first detected photon, calculated from the photon timestamps estimated using our proposed method. Representative results for photon index corresponding to the 2nd, 8th, 14th, and 20th detected photons are presented. While no difference in the peak positions is observed for the time difference between the second and first photons, the distributions for subsequent photons exhibit a temporal separation between fast and slow events. Based on these distributions, the classification accuracy for fast events was derived by progressively varying the timing threshold used for event separation (Fig. 5b). The results indicate that higher classification accuracy is achieved when the photon index difference is large and when a lower threshold is applied. Fig. 5c shows the fraction of events predicted as fast or slow under the same threshold conditions, and Fig. 5d provides the corresponding evolution of the FWHM and full width at tenth maximum (FWTM) of the time-difference histograms. The CTR was calculated from the time-difference histogram obtained after excluding events below a threshold determined from the estimated photon timestamps, based on the histogram shown in Fig. 5e (left). Although lower threshold values yield higher classification accuracy for fast events, they also reduce the survival fraction or detection efficiency of fast events. For example, when event separation is performed using the photon-index difference between the 20th and first detected photons estimated by the proposed method, setting the threshold to 141 ps leads to improvements of 16% in FWHM and 18% in FWTM; however, the number of retained fast events is substantially reduced (Fig. 5e, right histogram). In contrast, when the threshold is instead set to 423 ps, moderate improvements of 5% in FWHM and 6% in FWTM are achieved, while both the survival fraction and the classification accuracy of fast events are simultaneously maintained at approximately 60 % (Fig. 5e, center histogram). As the EMG-based slow tail which contributes to FWTM is due to scintillation-triggered events, the consequent FWTM enhancement is attributed to discrimination of Cherenkov and scintillation events.

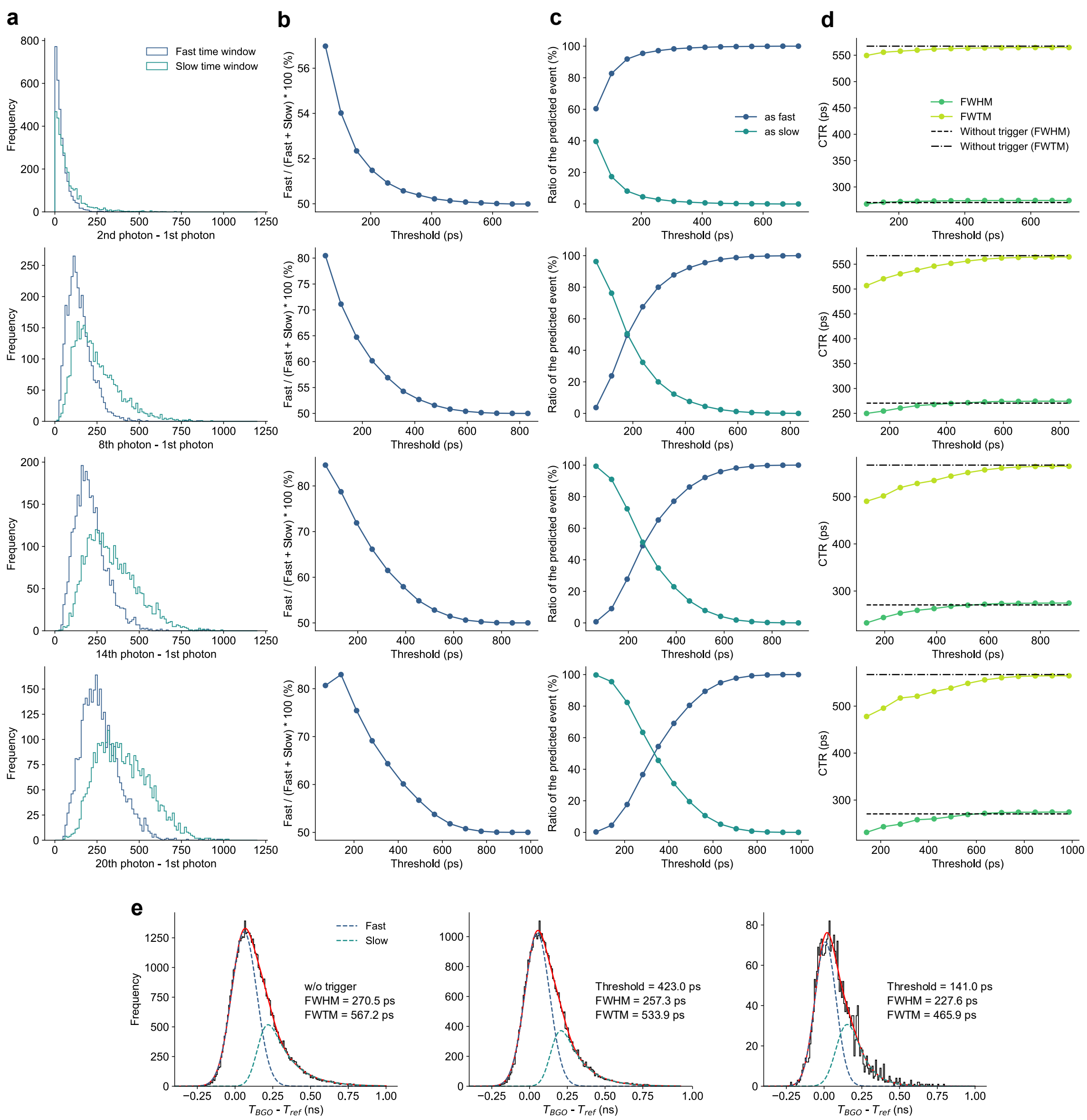

**Fig. 5 | Separation of Cherenkov- and scintillation-dominated events in BGO emission using estimated photon timestamps. a**, Time difference distributions between a specific photon and the first detected photon, calculated from photon timestamps estimated by the proposed method, shown separately for fast and slow events. Representative results are shown for the 2nd, 8th, 14th, and 20th photon-indexes from upper to bottom. **b**, Classification accuracy for identifying fast events as a function of the threshold applied to the photon time-difference distributions. **c**, Ratio of events classified as fast or slow under the same threshold conditions, illustrating the trade-off between classification accuracy and the survival ratio of fast events. **d**, FWHMs and FWTMs of the time-difference histograms before and after event separation. **e**, Time-difference histograms obtained from coincidence measurements between a BGO detector and a LYSO reference detector. CTR improvement by event separation using the 20th photon-index with different threshold values (center and right).

## Discussion

This study demonstrated that photon-level temporal information, long considered experimentally inaccessible in scintillation detectors, can be estimated and systematically analyzed using physics-informed unsupervised learning. By transforming conventional detector waveforms into a photon-resolved representation, our proposed approach reveals previously hidden structure within scintillation signals. This approach enables direct experimental validation of photon-level timing, thereby facilitating progress in a research area that has long remained confined to theoretical and simulation-based studies. To the best of our knowledge, the proposed method provides the first experimental demonstration that recovers all detected photon timestamps in a standard detector configuration, consisting of a one-to-one coupling between a scintillator and an SiPM, without any modification to the detector structure. Experimental evaluations demonstrated that CTR behavior (Fig. 4a) calculated using the estimated timestamps were consistent with those reported in previous theoretical and simulation-based studies[11-14] including all real parameters that are impossible to be parameterized in simulations. This finding establishes a new experimental capability for probing the internal temporal structure of scintillation events.

Beyond timing performance, the present results reveal an additional potential application enabled by the estimated photon-level information. Monte Carlo simulation showed that the proposed method preserves not only timing information but also photon counts with high accuracy (Fig. 2d). Accurate count preservation is essential for energy estimation[31]. In PET, reliable energy information is a critical requirement for event selection and scatter rejection, thereby enhancing image contrast and quantitative accuracy. The ability to estimate both photon timing and count therefore suggests that the proposed method is not limited to timing optimization, but instead provides a unified photon-level description of detector signals. Furthermore, the DOI analysis highlights the significance of photon timestamps. Using estimated photon arrival times, we experimentally confirmed that photon timing distributions vary as a function of interaction depth within the scintillator (Fig. 4d). The ability to access this information indicates that photon timestamps may provide direct sensitivity to spatial information of interaction between gamma rays and scintillators. This finding suggests that DOI information is inherently encoded in the photon-level temporal structure of scintillation signals, even when conventional timing estimators fail to capture it[17]. Importantly, this capability opens a pathway toward photon-based DOI estimation and combined TOF-DOI estimation without requiring additional device modifications. In contrast, conventional TOF-DOI detector designs often rely on specialized geometries[32-34] or multi-ended readout[19,35] architectures to extract DOI information. Our proposed approach instead leverages information already present in conventional waveform data, but previously inaccessible, thereby enabling advanced detector designs and photon-recovery strategies through software-based signal interpretation alone.

In BGO measurements, the estimated photon timestamps provided a mechanism for separating Cherenkov and scintillation photons, enabling both experimental and quantitative analyses of their respective contributions (Fig. 5). Event separation approaches based on analog waveform features, such as rise-time analysis[36], have been investigated by Kratochwil *et al.*. Our present results provide additional support for their work from the perspective of explicit photon-level timing. By resolving individual photon arrivals, the proposed method offers a physically grounded explanation for waveform-based separation phenomena. Threshold-based selection using estimated two-photon information revealed a trade-off between fast-event classification accuracy and detection efficiency. The optimal operating point must balance timing resolution and detection efficiency to maximize signal-to-noise performance. It should be noted that, from a system perspective, slow events are not necessarily discarded and can still contribute to image reconstruction[37], such that the overall statistical sensitivity is not inherently reduced. In this study, a CTR condition maintaining the accuracy and the survival fraction at about 60% for each was selected as a representative operating point. However, systematic modeling of this trade-off and principled optimization of selection parameters depend on applications and remain as future work.

Although the observed improvement in timing resolution was approximately 10%, such an enhancement is non-negligible in the context of ultrafast timing regimes. Achieving a 10% improvement in CTR through hardware alone would require, when focusing on the photon detection efficiency (PDE) in SiPM, a substantial relative increase of approximately 23%[38]. Considering recent hardware advancements[38-41], this represents a significant technical barrier; therefore, even such gains as 10% represent a meaningful step forward. In TOF-PET scanners, even modest improvements can enhance the signal-to-noise ratio, contributing to diagnostic capabilities. Furthermore, achieving higher levels of timing resolution would enable more precise event localization, which could facilitate flexible detector geometries and reconstruction-free functional[42] and anatomical[43,44] imaging paradigms, expanding diagnostic capabilities beyond conventional frameworks.

Device-based solutions such as digital SiPMs are capable of providing photon-level information directly. However, these devices have not yet been realized for radiation measurement applications and only simulation results can be discussed. Moreover, digital SiPMs typically involve high development cost and complexity, and the requirement to store all detected photon information places substantial demands on memory bandwidth and power consumption. Although single-photon avalanche diode-based cameras with more than a megapixel sensor resolution have been developed[45], their applications are concentrated on time-gated TOF measurement. For general-purpose TOF-PET systems, such designs may be considered over-specified. To overcome the issue, intermediate architectures between digital and analog SiPMs such as μSiPM[46], OctaSiPM[47], and semi-monolithic configuration[48] have been developed to capture limited photon timing information by segmenting the SiPM array coupled to a single scintillation crystal. These approaches partially mitigate hardware constraints, although they still require specialized detector designs and remain limited in the number of photons that can be effectively recorded. Moreover, those methods cannot be applied to scintillator crystals smaller than the photodetectors, limiting flexibility. The Digital Photon Counter developed by Philips enables fully digital readout; however, only a single timestamp is available per detector pixel of approximately $4^2$ $mm^2$ in size[49]. Therefore, the Digital Photon Counter retains a level of timing information comparable to that of conventional analog-based signal processing, for example, by a leading-edge discriminator or a constant fraction discriminator. In contrast, our proposed method operates entirely at the software level and enables photon-level information extraction without modifying the existing detector configuration, thereby enabling an analog SiPM to effectively function as a virtually digital SiPM. Because the timestamp estimation is performed from digitized waveforms, the number of photons to be extracted can be selectively adjusted according to application requirements or computational constraints. From this perspective, our method can be regarded as a cost-effective and flexible. We acknowledge that its applicability to PMT-based detectors remains unknown and should be investigated from the viewpoint of the stability of single photon response of PMTs compared to that of SiPMs.

Deconvolution techniques have been employed in detector signal processing[50-52]. However, single-photon response-based deconvolution studies have focused on estimating continuous scintillation probability density functions (PDFs) rather than directly estimating discrete photon timestamps. When inferring photon timestamps from continuous PDFs, the position of the first photon often must be manually initialized, which introduces significant uncertainty in applications where early photons dominate timing performance. The proposed framework adopts a straight-through estimator (STE)[53] structure in the network output layer, enabling binary outputs while preserving gradient backpropagation. This design allows direct and automatic estimation of photon timestamps including the first-arriving photon without manual intervention. More broadly, this work illustrates the potential of physics-informed unsupervised learning as a general paradigm for detector signal interpretation. From a methodological perspective, our proposed approach can be regarded as an extension of neural network-based frameworks combined with forward projection models, as commonly employed in image reconstruction[54], to detector signal analysis. A key aspect of this approach is that the high representational capacity of the neural network is coupled with an explicit physics-based forward model of the detector response and regularization related to photon dynamics, preventing convergence to solutions that merely fit noise. Recent advances in image reconstruction field have shown that augmenting unsupervised learning with explicit forward models[55] and task-specific regularization[56,57] substantially improves robustness and interpretability. By tightly coupling detector physics with optimization-based inference, the proposed approach achieves stable, interpretable, and adaptable photon timestamps estimation under realistic experimental conditions.

Despite these advantages, we should acknowledge several limitations of the proposed approach. First, it does not provide completely accurate photon timestamps, at least in its current form. Although the estimation errors are generally small, simulation results indicate that they are more pronounced for the earliest photons (Fig. 2c), which play a dominant role in timing performance. Accordingly, the results reported in this study, particularly those related to the first-arriving photons, should be interpreted with awareness of the underlying estimation uncertainty. In addition, the achievable timing performance is fundamentally limited by the single photon time resolution (SPTR) of the SiPM. Since SPTR introduces intrinsic temporal jitter at the level of individual detected photons, this effect is especially critical for early photons, which dominate timing resolution and are therefore most sensitive to SPTR-induced fluctuations. Likewise, optical crosstalk in SiPMs cannot be neglected. Therefore, further improvements in photon-level timing reconstruction require not only algorithmic advances but also photodetector technology advances. The estimation performance is also sensitive to hyperparameters of the network. In experimental validation, the plausibility and consistency of the obtained results must therefore be assessed comprehensively rather than relying on individual metrics alone. Reducing the dependence on hyperparameter selection and developing more parameter-robust models remain important directions for future improvement. The second limitation of our approach is that the event-by-event optimization required for timestamps estimation introduces additional computational cost compared with conventional timing estimation methods. Third, because the proposed method relies on digitized waveform data, its applicability is inherently limited to detector readout. Further developments such as algorithmic acceleration or hardware-aware optimization will be necessary to enable real-time processing or deployment in high-count-rate systems.

On the other hand, in applications of scintillation detectors in fields other than PET, these limitations can often be mitigated. In many neutrino experiments, the waveforms of all PMTs are fully digitized using flash analog-to-digital converters, as these experiments aim to search for extremely rare physical events and the data acquisition systems are typically designed to extract as much information as possible from the digitized waveforms[58,59]. High-energy physics and neutrino experiments also demand extremely precise timing performance similar to PET. Precise timing enables effective background rejection, separation of Cherenkov and scintillation light components, and improved particle identification, particularly in environments with high event rates or low signal-to-noise ratios[60-63]. Applying the proposed approach is expected to enhance spatiotemporal information and improve event identification. Such capabilities could ultimately contribute to improved experimental precision and deeper insights into fundamental physics.

# Methods

### Monte Carlo simulation setup

The detector was modeled using Geant4 (ver. 11.0) as a combination of an LYSO crystal, reflectors, an optical coupling layer, and an SiPM. To evaluate the validity of the simulation setup using the CTR metric, two identical detectors were placed face-to-face, with an annihilation gamma-ray source positioned at the center between them. The crystal dimensions were set to 3 × 3 × 10 $mm^3$. The optical properties were defined as a refractive index of 1.8 and a scintillation decay time of 37.1 ns[34]. The absorption and Rayleigh scattering lengths were implemented as wavelength-dependent parameters[6]. An optical grease having $SiO_2$, as its base component was applied as a layer between the crystal and the SiPM, while aluminum reflectors were applied to the lateral and rear surfaces of the crystal. The optical interfaces between the crystal and surrounding materials were modeled using the LUTDavis model[64] to simulate realistic polished crystal surfaces. The generated optical photons propagated through the crystal while undergoing scattering, absorption, and boundary reflections before reaching the SiPM surface. The SiPM was modeled as a photosensitive volume of 3 × 3 × 0.1 mm$^3$, and an optical photon was defined as being detected when it was absorbed within the SiPM volume. For each event, the arrival time and wavelength of detected photons, as well as the number of detected photons, were recorded in list-mode format.

Detected photons were sampled according to the PDE of the SiPM obtained from the manufacturer's datasheet during offline signal processing. To account for the SPTR, a time jitter modeled as a Gaussian distribution with 140 ps FWHM was added to each photon timestamp[40]. Subsequently, the arrival times of scintillation photons were sorted chronologically, and each photon timestamp was convolved with a measured single-photon response function to generate analog waveforms. Both the photon timestamps and the reconstructed waveforms were generated with a sampling interval of 1 ps. Amplitude fluctuations and baseline electronic noise were modeled as Gaussian distributions based on experimentally measured single-photon response. As a consistency check, the CTR was evaluated using a conventional timing pickoff approach based on pulse amplitude. By varying the timing threshold from values close to the baseline to higher fractions of the pulse amplitude, the resulting CTR values were as 135.5 ps, 131.8 ps, 133.6 ps, 135.3 ps, and 137.8 ps. Although, it should be emphasized that the purpose of this signal simulation was not to achieve a highly detailed or fully realistic reproduction of the analog response, these results confirm that the reconstructed waveforms yield timing performance generally consistent with that of typical experimental detector systems.

**Experimental setup for CTR evaluation**

Fig. 3 illustrates the experimental setup for the acquisition of coincidence waveforms. Each detector consisted of a multi-pixel photon counter[65] (MPPC; Hamamatsu Photonics K.K., Japan) with an active area of 4 × 4 mm$^2$ coupled to a 3 × 3 × 10 mm$^3$ LYSO crystal (EPIC Crystal, China) that had been wrapped with an enhanced specular reflector and subsequently covered with Teflon tape. Optical coupling between the crystal and MPPC was conducted using Meltmount ($n$ = 1.58). A $^{22}$Na point source was positioned at the center between the two detectors.

For coincidence measurements, the readout of the output signals, provided via high-frequency electronics, were fully digitized using an oscilloscope (DSO-404A; Keysight Technologies, USA) operating at a sampling rate of 20 GS/s with a bandwidth of 4.2 GHz. To accurately capture the leading edges of the timing signals from both detectors, the vertical range of the oscilloscope was intentionally restricted. A total of 400 sampling points were recorded for timing signals. Energy signals were employed exclusively for oscilloscope triggering and were neither displayed nor stored to preserve the maximum achievable sampling rate. The energy threshold was determined from the pulse-height distribution of the energy signals and set near the valley between the photopeak and the Compton edge, approximately corresponding to 420 keV. The CTR was calculated by fitting a Gaussian function to the time-difference histogram between two detectors, with the value at FWHM selected as the final CTR.

**Experimental setup for DOI evaluation**

One of the two detectors was identical to the reference detector used in the CTR measurements, while the other was modified for DOI evaluation by replacing its scintillator crystal with a 3 × 3 × 20 $mm^3$ LYSO crystal (EPIC Crystal, China). The target detector was arranged with its long axis oriented perpendicular to that of the reference detector (Fig. 3). A lead block was employed for collimation to selectively control the gamma-ray interaction position within the target detector. By adjusting the collimation geometry, coincidence events corresponding to interactions occurring near the SiPM-coupled end (referred to as the bottom position) and those occurring farther from the SiPM (referred to as the top position) were individually acquired. Coincidence data from these two interaction positions were collected and analyzed under the same experimental conditions as those used in the CTR evaluation.

**Experimental setup for BGO evaluation**

One of the detectors was equipped with a BGO crystal (EPIC Crystal, China) having dimensions of 3 × 3 × 5 $mm^3$, while the opposing detector was the LYSO-based reference detector described for the CTR evaluation. All other experimental conditions were identical to those used in the CTR evaluations described above. The obtained time-difference histogram exhibited an asymmetric distribution with a longer tail on one side. This distribution arises from the coexistence of two distinct emission components in BGO: a prompt component mainly attributed to Cherenkov photons and a delayed component arising from scintillation photons. The time-difference distribution was fitted with a composite function consisting of a Gaussian component and an EMG component, as follows:

$$f(t) = A_G \exp\left(-\frac{(t-\mu_G)^2}{2\sigma_G^2}\right) + \frac{A_E}{2\tau}\exp\left(\frac{\mu_E}{\tau} + \frac{\sigma_E^2}{2\tau^2} - \frac{t}{\tau}\right)\mathrm{erfc}\left(\frac{\mu_E + \sigma_E^2/\tau - t}{\sqrt{2}\sigma_E}\right), \mathrm{erfc}(t) = \int_t^\infty e^{-\lambda^2} d\lambda, (1)$$

where $A$, $\mu$, $\sigma$, and $\tau$ are amplitude, mean, standard deviation of the Gaussian component, and decay time of the exponential component, respectively. The former component represents the contribution of prompt Cherenkov events, whereas the latter component accounts for the slower tail due to the scintillation-events contribution.

**Measurement of single-photon response**

The single-photon response of the SiPM was measured using high-frequency electronics and a pulsed laser with a pulse width of 60 ps FWHM (C10196; Hamamatsu Photonics K.K., Japan)[51]. The laser intensity was attenuated to the single-photon level using a neutral density filter. The SiPM was biased at the same applied voltage as that used in the experimental setups described above, such that the single-photon signal amplitude was sufficiently larger than the electronic noise level. The output signal from the SiPM was amplified and fed into an oscilloscope, where waveforms were recorded on an event-by-event basis. To guarantee the triggering time of the SiPM signals, the oscilloscope was triggered by a synchronized laser signal. The maximum amplitude of each recorded waveform event-by-event was used to construct a pulse-height histogram, and single-photon equivalent events were selected based on this histogram by isolating the first peak. The selected waveforms were baseline-corrected and averaged to obtain the representative single-photon response waveform.

**Deep learning-based deconvolution**

Fig. 1b depicts an overview of the proposed method. For each acquired event, a neural network maps a detector waveform to quantized photon timestamps that corresponds to a binary temporal sequence indicating the occurrence of detected photons. The training process of this network $\boldsymbol{f_\theta}$ is optimized independently for each event by minimizing an objective function consisting of a waveform reconstruction loss and a regularization term without ground-truth timestamp information:

$$\boldsymbol{\theta}^* = \underset{\theta}{\operatorname{argmin}} \|\boldsymbol{S} \cdot \boldsymbol{f_\theta}(\boldsymbol{z}) - \boldsymbol{w}\| + \lambda \sum_k \alpha_k \cdot f_\theta(z)_k, \qquad \boldsymbol{t}^* = \boldsymbol{f}_{\boldsymbol{\theta}^*}(\boldsymbol{z}), \tag{2}$$

where $\|\cdot\|$ is the L2 loss, $\boldsymbol{z}$ is a constant input vector, $\lambda$ is a regularization weight, and $\alpha_k$ represents a time-dependent weighting coefficient applied to each discrete time bin $k$. In the first term of Eq. (2), the output timestamps are convolved with the pre-acquired single-photon response of the SiPM to reproduce the detector physical model within the deep learning framework, and the error between the reconstructed waveform and the measured detector waveform $\boldsymbol{w}$ is calculated. To compute waveform reconstruction in a fully differentiable manner, the convolution operation is reformulated as a matrix-vector multiplication. Specifically, the single-photon response is converted into the following matrix $\boldsymbol{S}$, constructed from shifted versions of the single-photon response:

$$\boldsymbol{S} = \begin{bmatrix} v(t_1) & v(t_2) & \cdots & v(t_N) \\ 0 & v(t_1) & \cdots & v(t_{N-1}) \\ \vdots & \vdots & \ddots & \vdots \\ 0 & 0 & \cdots & v(t_1) \end{bmatrix}, \tag{3}$$

where $v$ denotes the amplitude at each time bin from $t_1$ to $t_{1-N}$ of the single-photon response, and the first row of this matrix corresponds to the original response waveform. The reconstructed waveform is given by $\boldsymbol{S} \cdot \boldsymbol{f_\theta}(\boldsymbol{z})$. The time-weighted L1 regularization for output timestamps imposes a penalty that adjusts the number of photons to suppress the influence of noise, while placing emphasis on consistency at earlier photons that contain more informative timing information. Through iterative optimization, the network output produces the final photon timestamp sequence $t^*$.

**Implementation details**

We implemented the network as a four-layer multilayer perceptron. Each hidden layer consists of 100 nodes with ReLU activations. The output layer is designed to produce binary output corresponding to photon arrival events on discrete time bins. To enable gradient-based optimization with discrete outputs, we introduced the STE[53] structure. The output layer produces continuous values in [0,1] using a sigmoid activation; these are then binarized using a threshold of 0.5 during the forward pass. In the backward pass, the non-differentiable binarization operation is approximated by the identity function, allowing gradients to pass through the output layer unchanged. Each output time bin corresponds to a temporal resolution of 1 ps, matching the sampling interval of the waveform. Optimization was performed using AdamW with a learning rate of 0.01 and an iteration number of 800, and this condition was kept constant across all experiments. The time-dependent regularization weights were linearly varied over the full time window such that the earliest time bin was penalized twice as strongly as the latest one. The network model was built using PyTorch 1.6.0 and a graphics processing unit (TITAN RTX; NVIDIA, USA) with 24 GB capacity.

**Timing estimators using multiple timestamps**

To calculate the CTR from estimated multiple timestamps $\boldsymbol{t}$, an appropriate timing estimator is required. The simplest approach is to average the arrival times of multiple photons, which is defined as

$$x_{AVG} = \frac{1}{N}\sum_{i=1}^{N} t_i \,, \tag{4}$$

where $N$ denotes the number of timestamps used in the averaging. An alternative approach[13] is to estimate the most likely time $x$ by maximizing the probability density function $P(x|\boldsymbol{t})$. Bayes' theorem can be used to wrote the posterior probability density function in terms of the likelihood function $P(\boldsymbol{t}|x)$ as

$$P(x|\boldsymbol{t}) = \frac{P(x) \cdot P(\boldsymbol{t}|x)}{P(\boldsymbol{t})}. \tag{5}$$

In this study, the prior distribution $P(x) = 1$. The denominator acts as a normalization constant and does not depend on $x$. Assuming that the probability density function of each photon detection can be approximated by a Gaussian distribution, we can express $P(\boldsymbol{t}|x)$ as a multivariate normal distribution:

$$P(t|x) = \frac{1}{\sqrt{(2\pi)^N |\boldsymbol{\Sigma}|}} \exp\left[-\frac{1}{2}(\boldsymbol{t} - \boldsymbol{\mu} - \mathbf{1}x)^T \boldsymbol{\Sigma}^{-1} (\boldsymbol{t} - \boldsymbol{\mu} - \mathbf{1}x)\right], \tag{6}$$

where $\mathbf{\Sigma}^{-1}$ denotes the inverse of the covariance matrix, and each element of the vector $\boldsymbol{\mu}$ represents the centroid time of the corresponding photon emission. Maximizing $P(x|\boldsymbol{t})$ is equivalent to setting its derivative with respect to $x$ to zero, which corresponds to minimizing the exponent of the likelihood function. This represents the following maximum likelihood estimator:

$$x_{MLTE} = \frac{\mathbf{1}^T\mathbf{\Sigma}^{-1}}{\mathbf{1}^T\mathbf{\Sigma}^{-1}\mathbf{1}}(\boldsymbol{t} - \boldsymbol{\mu}). \tag{7}$$

The estimator $x_{MLTE}$ can be interpreted as a weighted average of the $N$ measured timestamps. The weighting factors depend solely on the calibration run, as encoded in the covariance matrix $\mathbf{\Sigma}$ and the mean vector $\boldsymbol{\mu}$. This time estimator can also be derived using the generalized least-squares method and is referred to as the Gauss-Markov estimator in Veniago *et al.*[14].

## Acknowledgments

This work was supported by the Nakatani Foundation. The authors are grateful to Mr. Takahiro Moriya from the Central Research Laboratory at Hamamatsu Photonics K. K. and the members of the Imaging Physics Group at National Institutes for Quantum Science and Technology for their kind support.